\documentclass{llncs}

\usepackage{amsmath,amsfonts, amssymb,tikz,xcolor}

\newcommand{\ZZ}{\mathbb{Z}}
\newcommand{\NN}{\mathbb{N}}

\newcommand{\p}{p}

\newcommand{\ACA}{\mathcal{A}}
\newcommand{\ACB}{\mathcal{B}}
\newcommand{\ACU}{\mathcal{U}}
\newcommand{\alphA}{{Q_\ACA}}

\newcommand{\alphU}{{Q_\ACU}}
\newcommand{\locA}{\delta_\ACA}

\newcommand{\csetA}{\alphA^\ZZ} 
 
\newcommand{\csetU}{\alphU^\ZZ}

\newcommand{\fmap}{\phi}
\newcommand{\fsimu}{\preccurlyeq}  

\newcommand{\mo}{\mathbb{M}}

\newcommand{\psta}{q_0}
\newcommand{\pset}{\mathbb{Pe}}
\newcommand{\pseto}{\mathbb{Pe}_0}

\newcommand\rect[1]{\mathcal{R}_{\vec{{#1}}}}
\newcommand\patternatx[3]{\mathcal{P}^{{#1}}_{{#2}}\left(#3\right)}
\newcommand\patternat[3]{\patternatx{\vec{#1}}{\vec{#2}}{#3}}

\newtheorem{defi}{Definition}
\newtheorem{prop}{Proposition}
\newtheorem{theo}{Theorem}
\newtheorem{lemm}{Lemma}
\newenvironment{proo}[1][]{\textit{Proof #1:}}{\qed }

\title{On Factor Universality in Symbolic Spaces}

\author{Laurent Boyer\thanks{\email{laurent.boyer@univ-savoie.fr}}\inst{1} and Guillaume
  Theyssier\thanks{\email{guillaume.theyssier@univ-savoie.fr}}\fnmsep\thanks{Partially supported by French ANR 'projet blanc' EMC (NT09\_555297)}\inst{1}}

\institute{LAMA, (UMR 5127 --- CNRS, Universit\'e de Savoie), Campus
  Scientifique,\\ 73376 Le Bourget-du-lac cedex FRANCE}

\begin{document}

\maketitle

\begin{abstract}
  The study of factoring relations between subshifts or cellular automata is
  central in symbolic dynamics. Besides, a notion of intrinsic universality for
  cellular automata based on an operation of rescaling is receiving more and
  more attention in the literature. In this paper, we propose to study the
  factoring relation up to rescalings, and ask for the existence of universal
  objects for that simulation relation.
  
  In classical simulations of a system $S$ by a system $T$, the simulation takes
  place on a specific subset of configurations of $T$ depending on $S$ (this is
  the case for intrinsic universality). Our setting, however, asks for every
  configurations of $T$ to have a meaningful interpretation in $S$. Despite this
  strong requirement, we show that there exists a cellular automaton able to
  simulate any other in a large class containing arbitrarily complex ones.  We
  also consider the case of subshifts and, using arguments from recursion theory, we
  give negative results about the existence of universal objects in some
  classes.
\end{abstract}

\section{Introduction and definitions}

Tilings and cellular automata are two paradigmatic models often
considered in the fields of complex systems and natural
computing. They are complementary ---one is static and
non-deterministic and the other is dynamic and deterministic--- but
they are both formally simple and both related to symbolic
spaces. Moreover, many links are now established between the two
models (see for instance \cite{kari92,HochmanEff}) so it is natural to
consider them together.

Both are known to be Turing-powerful since their introduction in the
mid-20th century \cite{Wang,vonneumann}. However, analyzing their
ability to process information only through translations into the
Turing world is very restrictive. Such models of natural computing
deserve a natural and intrinsic notion of reduction to compare their
objects one to each other. Following this line of thought, several
notions of simulations were proposed recently which are intrinsic to
each model, and lead to corresponding intrinsic notions of
universality \cite{ollingerphd,theyssierphd,LafitteWeiss,self-assembly}: a system
is universal if it is able to simulate any other from the same class.

Intrinsic universality for cellular automata is probably the most
studied of such notions
\cite{Ollinger,Ollinger03,Moreira03,BoyerT09,Durand-Lose97}. The
underlying relation of simulation uses uniform encodings working at
the level of blocks of cells. More precisely, S simulates T if S, when
restricted to a suitable subset of 'correct' configurations, is
isomorphic to T via such an encoding. Our approach is different and
uses redundancy of information instead of restriction to a subset of
configurations. In our setting, S simulates T if there is a uniform
way of projecting \emph{the whole} phase space of S onto the phase
space of T (a precise definition is given below). The question
addressed by this paper is the existence of universal objects with
respect to that simulation relation, we call them \emph{factor
  universal} objects. 

The first contribution of this paper is the formalism based on the
well-known mathematical notion of action: it allows to encompass both
subshifts and cellular automata, it gives a new look at the notion of
cell grouping which is the root of the simulation relation used in
intrinsic universality, and it establishes connections with the work of
Hochman \cite{HochmanUniv} where the use of sub-actions is
crucial. Our main result is that, although factor-universal objects do
not generally exist (theorem~\ref{theo:nouniv}), it can still be
constructed for some large class like the set of cellular automata
having a persistent state (theorem~\ref{theo:univpe}).

\paragraph{\textbf{Basic definitions.}}
Given a finite set $Q$ and an integer ${d\geq 1}$, the \emph{symbolic space} of
\emph{dimension} $d$ over \emph{alphabet} $Q$ is the set $Q^{\ZZ^d}$. It can be
seen as an infinite set of cells arranged as a lattice $\ZZ^d$ and each carrying
a value from $Q$. An element of $Q^{\ZZ^d}$ is called a
\emph{configuration}. $Q^{\ZZ^d}$ is naturally equipped with the compact Cantor
topology \cite{kurkabook} which is the product topology of the discrete topology
on $Q$ (it can also be defined via a metric). 

Another key notion in the context of symbolic spaces is that of \emph{finite
  patterns} that may occur in infinite configurations. For our purpose,
rectangular patterns will be enough.  Given $\vec{z}=(z_1,\cdots,z_d)\in\ZZ^d$
with $z_i> 0$ for all $i$, the hyperrectangle $\rect{z}$ is the set of vectors
$\vec{z'}=(z'_1,\cdots,z'_d)\in\ZZ^d$ such that ${0\leq z'_i< z_i}$ for all $i$.
A $Q$-pattern of shape $\rect{z}$ is a coloring of $\rect{z}$ by $Q$, that is an
element of $Q^{\rect{z}}$. Given a configuration $c\in Q^{\ZZ^d}$, the pattern
of shape $\rect{z_s}$ extracted from $c$ at position $\vec{z_p}\in\ZZ^d$,
denoted by $\patternat{z_s}{z_p}{c}$, is simply: ${z\in\rect{z_s}\mapsto c(z_p+z).}$

The objects we study (subshifts and cellular automata) share the property of
being uniform, \emph{i.e.} invariant by translations. Formally, given
${z\in\ZZ^d}$ the translation of vector $z$, denoted $\sigma_z$, is the function
mapping a configuration $c\in Q^{\ZZ^d}$ to the configuration $\sigma_z(c)$ such that
${\forall z'\in\ZZ^d}$, ${\sigma_z(c)({z'}) = c({z'+z})}$.

A \emph{subshift} is a subset of $Q^{\ZZ^d}$ which is translation invariant and
closed for the Cantor topology. Equivalently, a subshift is a set $\Sigma_L$ of
configurations avoiding any occurrence of any finite pattern from a given
language of patterns $L$:
\[\Sigma_L = \bigl\{c\in Q^{\ZZ^d}:\ \forall z,z'\in\ZZ^d \text{ with
  }z_i>0\text{ for all }i,\
\patternat{z}{z'}{c}\not\in L\bigr\}.\]

A subshift of \emph{finite type} is a subshift of the form $\Sigma_L$ where $L$
is finite. There are strong connections between subshifts of finite type in
dimension $2$ and sets of tilings generated by a set of wang tiles. In
particular, due to Berger's theorem \cite{berger}, it is undecidable, given a
finite $L$, to determine whether $\Sigma_L$ is empty or not.

A \emph{cellular automaton} is a local and uniform map on a symbolic
space. Formally, it is given as a $4$-tuple by its dimension $d$, its alphabet
$Q$, its \emph{neighborhood} ${V\subseteq \ZZ^d}$ (finite) and its \emph{local
  transition map} ${f: Q^V\rightarrow Q}$. To that formal object we associate a
\emph{global map} $F$ acting on ${Q^{\ZZ^d}}$ as follows:
\[\forall c\in\ZZ^d, \forall z\in\ZZ^d,\ F(c)(z) = f\bigl(z'\in V\mapsto
c(z+z')\bigr).\] The fundamental theorem of Curtis-Lyndon-Hedlund
\cite{hedlund} states that global maps of cellular automata are exactly
continuous maps on symbolic spaces which commute with translations.

\paragraph{\textbf{Actions and rescalings.}}
Let $(\mo,+)$ be a monoid (a set equipped with an associative law and
a neutral element). An $\mo$-\emph{action} on a space $X$ is a
function $\Psi :\mo\times X\rightarrow X$ such that ${\Psi(0,x) = x}$
(for all $x\in X$ and $0$ being the neutral element of $\mo$) and
\[\forall x\in X, \forall m,m'\in\mo,\ \Psi(m+m',x) = \Psi\bigl(m,\Psi(m',x)\bigr).\]
We will use the formalism of action to study both subshifts and cellular automata:
\begin{itemize}
\item if ${\Sigma\subseteq Q^{\ZZ^d}}$ is a subshift, we canonically
  associate to it the $\ZZ^d$-action ${\Psi}_\Sigma$ on $\Sigma$
  defined by ${{\Psi}_\Sigma(z,x) = \sigma_z(x)}$;
\item if $F$ is a cellular automaton on the space $Q^{\ZZ^d}$, we
  canonically associate to it the $\NN\times\ZZ^d$-action ${\Psi}_F$
  on $Q^{\ZZ^d}$ defined by ${{\Psi}_F\bigl((t,z),x\bigr) = \sigma_z\circ F^t(x)}$.
\end{itemize}

If $\mo'$ is a sub-monoid of $\mo$, $\Psi$ induces an $\mo'$-action by
restriction to the domain $\mo'\times X$. $\mo$ and $\mo'$ can be isomorphic or
not and both cases might be interesting. For instance, studying a cellular
automaton $F$ as a classical dynamical system consists in forgetting the spacial
component of $\Psi_F$ and focusing on the pure temporal action of $F$. This
point of view was often adopted in the literature (\emph{e.g.}, topological
dynamics of cellular automata \cite{kurkabook}) but, interestingly enough,
recent work of Sablik \cite{Sablik08} tends to re-incorporate the spacial
component of actions to better study the dynamics of cellular automata.

In this paper, we will only consider the case where $\mo$ and $\mo'$
are isomorphic. More precisely, in our context, $\mo$ will be of the
form $\ZZ^d$ or $\NN\times\ZZ^d$ and we will consider sub-monoids of
the form ${\mo' = t_0\NN\times z_1\ZZ\times\cdots\times z_d\ZZ}$, with
$t_0>0$ and $z_i>0$ for all $i$.  In this case, passing from the
$\mo$-action to the $\mo'$-action can be seen as a neutral change of
point of view on the system that we call \emph{rescaling} in the
sequel. The intuition is that we change the discrete units of time and
space, passing from $1$ to $t_0$ in time and $1$ to $z_i$ in direction
$i$. Given a subshift or a cellular automaton, a \emph{scaled action}
is simply the restriction of their canonical action to some sub-monoid
of the form $\mo'$. It is worth noticing that a scaled action
associated to a subshift (resp. a cellular automaton) on the alphabet
$Q$ is always isomorphic to the canonical action of a subshift
(resp. a cellular automaton) on an alphabet of the form $Q^k$. More
concretely, this isomorphism comes from the natural one-to-one map
from $Q^{\ZZ^d}$ to ${\left(Q^{\rect{z_s}}\right)^{\ZZ^d}}$, where
${\vec{z_s}=(z_1,\ldots,z_d)}$, which maps a configuration $c$ to: ${z
  \mapsto \patternat{z_s}{z\times z_s}{c}}$, where the operation
$\times$ on $\ZZ^d$ denotes coordinate-wise multiplication. Our notion
of rescaling for cellular automata is similar to the one in
\cite{ollingerphd,theyssierphd} which is the basic ingredient to
define intrinsic universality.

\paragraph{\textbf{Factors.}}
One of the central notion in symbolic dynamics is that of \emph{factor}.
Intuitively, a factor is a uniform continuous projection.  This notion has also
been used with success in the study of expansive cellular automata \cite{Nasu}
and more generally as a classification tools for cellular automata \cite{Kurka97,Guillon}.  As we study both multi-dimensional subshifts and cellular automata,
we give a unified definition using the formalism of actions.
\begin{defi}
  Let $\mo$ and $\mo'$ be isomorphic monoids via $i:\mo\rightarrow \mo'$.  We
  say an $\mo'$-action $\phi'$ on $X'$ is a \emph{factor} of a $\mo$-action
  $\phi$ on $X$ if there is a continuous onto map $\pi : X\rightarrow X'$ such
  that: ${\forall x\in X, \forall m\in \mo,\ \pi\bigl(\phi(m,x)\bigr) =
  \phi'\bigl(i(m),\pi(x)\bigr)}$.
\end{defi}
Two key points are that: (1) {any} orbit in $(\phi,X)$ projects onto {some}
orbit of $(\phi',X')$ via $\pi$, and (2) {any} orbit of $\phi'$ can be realized
as such a projection. In a word, the simulation of $(\phi',X')$ by $(\phi,X)$ is
\emph{everywhere meaningful} and \emph{complete}.

\section{Factor Universality}
\label{sec:univdef}

At this point, we could compare subshifts or cellular automata through the
factoring relation between their canonical actions, saying that system $S$
factors onto system $T$ if the canonical action of $S$ factors onto that of $T$.
However, this gives an excessive importance to the alphabet and forbid the
existence of universal objects due to entropy considerations (factoring cannot
increase entropy).
In \cite{HochmanUniv}, this limitation is bypassed via dimension
changes: a $d$-dimensional system is compared to $k$-dimensional
systems ($k<d$) via its $k$-dimensional sub-actions. Our point of view
is different. We always work at constant dimension, but we use another
kind of sub-actions: \emph{scaled actions} defined above. For a fixed
dimension, monoids of scaled actions are all isomorphic and we will
consider only canonical component-wise isomorphisms between them. We
can now formulate the central definition of the paper.

\begin{defi}
  Let $S$ and $T$ be two $d$-dimensional subshifts (resp. CA). We say
  that $T$ is simulated by $S$, denoted $T\fsimu S$, if some scaled
  action of $S$ factors onto some scaled action of $T$.
\end{defi}

As usual when working on symbolic spaces, continuity and uniformity
implies locality (Curtis-Lyndon-Hedlund theorem \cite{hedlund}).  In
our context of rescalings, the locality is no longer expressed at the
level of cells, but at the level of groups of cells. More precisely,
we say that a map ${\phi : Q_1^{\ZZ^d}\rightarrow Q_2^{\ZZ^d}}$ is
\emph{local} if there exist: ${r\in\NN}$ (locality radius), two shapes
$\rect{z_1}$ and $\rect{z_2}$ (source and destination scales), and a
local function $f : Q_1^{\rect{(2r+1)z_1}}\rightarrow
Q_2^{\rect{z_2}}$ such that
\[\forall c\in Q_1^{\ZZ^d},\forall z\in\ZZ^d,\ \patternat{z_2}{z\times
  z_2}{\phi(c)}=f\bigl(\patternatx{(2r+1)\vec{z_1}}{\vec{z\times z_1 -}r\vec{z_2}}{c}\bigr).\]

To fix ideas, if ${d=z_1=z_2=1}$ and $Q_1=Q_2$, $f$ is just the local map of a
cellular automaton of radius $r$ and $\phi$ is its corresponding global map.

\begin{prop}
  \label{prop:hedlund}
  Fix a dimension $d$. Let $\Sigma_1$ and $\Sigma_2$ be two $d$-dimensional
  subshifts and let $F_1$ and $F_2$ be two $d$-dimensional CA of alphabet $Q_1$
  and $Q_2$ respectively. Then we have:
  \begin{itemize}
  \item $\Sigma_2\fsimu\Sigma_1$ if and only if there is a local map $\phi$ such that 
    $\phi(\Sigma_1)=\Sigma_2$;
  \item $F_2\fsimu F_1$ if and only if there is an onto local map $\phi$ from
    $Q_1^{\ZZ^d}$ to $Q_2^{\ZZ^d}$ and integers $t_1,t_2\in\NN$ such that
    ${\phi\circ F_1^{t_1} = F_2^{t_2}\circ\phi}$.
  \end{itemize}
\end{prop}

Besides the work of Hochman \cite{HochmanUniv}, notions of simulations similar
to $\fsimu$ have already been considered for tilings \cite{LafitteWeiss} or for
cellular automata \cite{theyssierphd,bulking2} or more general settings
\cite{Kurka99}. Each time, one of the main concern is the existence of universal
objects: this is precisely the central point of the present paper.






\begin{defi}
  Let $\mathcal{C}$ be a class of subshifts (resp. cellular automata). A
  subshift (resp. cellular automaton) $U$ is \emph{$\mathcal{C}$-universal} if
  $U\in\mathcal{C}$ and $X\fsimu U$ for any $X\in\mathcal{C}$.
\end{defi}

Whatever the fixed dimension, there is no universal subshift for cardinality
reasons: there are uncountably many subshifts but for a given subshift $U$ there
are at most countably many different subshifts $\fsimu$-simulated by $U$ (by
proposition~\ref{prop:hedlund}). The following theorem uses recursion theoretic
arguments to yield other negative results concerning universality (similar
arguments where used in \cite{ollingerphd,HochmanUniv} in different settings).

\begin{theo}
  \label{theo:nouniv}
  Fix a dimension $d\geq 2$. Then there is no universal subshift of
  finite type of dimension $d$ and there is no surjective-universal CA
  of dimension $d$.
\end{theo}

\section{A Large Class with a Universal Object}
\label{sec:univpe}
In this section, we restrict to dimension 1 to make a clear exposition
of the main result (theorem~\ref{theo:univpe}).

\begin{defi}
A CA $\ACA$ is said to be \emph{persistent} if there is a state $q_0\in\alphA$ such that for any configuration $c\in \csetA$ if $c(i)= \psta$ then $\ACA(c)\ (i) =\psta$.

We denote by $\pset$ the set of all persistent CA.
\end{defi}

Note that for any CA, you may add an extra persistent state and obtain
a CA in $\pset$ containing the dynamics of the first one. 


\begin{theo}
  \label{theo:univpe}
  There exists a $\pset$-universal cellular automaton.
\end{theo}

\newcommand{\coA}{\mathcal{C}_\ACA}

\def\bsize{0.3cm}
\newcommand{\sun}[2]{	\node at (#1,#2) [rectangle,draw=black,fill=blue!20,inner sep=0pt,minimum size=\bsize] {$s_1$};}
\newcommand{\sunp}[2]{	\node at (#1,#2) [rectangle,draw=black,fill=blue!20,inner sep=0pt,minimum size=\bsize] {$s_1'$};}
\newcommand{\suns}[2]{	\node at (#1,#2) [rectangle,draw=black,fill=blue!20,inner sep=0pt,minimum size=\bsize] {$s_1''$};}
\newcommand{\sdeux}[2]{
	\node at (#1,#2) [rectangle,draw=black,fill=red!20,inner sep=0pt,minimum size=\bsize] {$s_2$};
	\node at (#1,#2+1) [rectangle,draw=black,fill=red!15,inner sep=0pt,minimum size=\bsize] {$s_2$};
	\node at (#1,#2+2) [rectangle,draw=black,fill=red!10,inner sep=0pt,minimum size=\bsize] {$s_2$};
}
\newcommand{\strois}[2]{
	\node at (#1,#2) [rectangle,draw=black,fill=green!20,inner sep=0pt,minimum size=\bsize] {$s_3$};
	\node at (#1,#2+1) [rectangle,draw=black,fill=green!15,inner sep=0pt,minimum size=\bsize] {$s_3$};
	\node at (#1,#2+2) [rectangle,draw=black,fill=green!10,inner sep=0pt,minimum size=\bsize] {$s_3$};
}
\newcommand{\squatre}[2]{	\node at (#1,#2) [rectangle,draw=black,fill=red!20,inner sep=0pt,minimum size=\bsize] {$s_4$};}
\newcommand{\squatrep}[2]{	\node at (#1,#2) [rectangle,draw=black,fill=red!20,inner sep=0pt,minimum size=\bsize] {$s_4'$};}
\newcommand{\scinq}[2]{	\node at (#1,#2) [rectangle,draw=black,fill=blue!20,inner sep=0pt,minimum size=\bsize] {$s_5$};}
\newcommand{\ssix}[2]{
	\node at (#1,#2) [rectangle,draw=black,fill=green!20,inner sep=0pt,minimum size=\bsize] {$s_6$};
	\node at (#1,#2+1) [rectangle,draw=black,fill=green!15,inner sep=0pt,minimum size=\bsize] {$s_6$};
	\node at (#1,#2+2) [rectangle,draw=black,fill=green!10,inner sep=0pt,minimum size=\bsize] {$s_6$};
}
\newcommand{\sr}[2]{	\node at (#1,#2) [rectangle,draw=black,fill=yellow!,inner sep=0pt,minimum size=\bsize] {$s_r$};}
\newcommand{\srp}[2]{	\node at (#1,#2) [rectangle,draw=black,fill=yellow!20,inner sep=0pt,minimum size=\bsize] {$s_r'$};}

\newcommand{\sM}[2]{	\node at (#1,#2) [rectangle,draw=black,fill=red!40,inner sep=0pt,minimum size=\bsize] {$M$};}
\newcommand{\dieze}[2]{	\node at (#1,#2) [rectangle,draw=black,fill=black!40,inner sep=0pt,minimum size=\bsize] {$\#$};}
\newcommand{\czero}[2]{	\node at (#1,#2) [rectangle,inner sep=0pt,minimum size=\bsize] {$C_0$};}
\newcommand{\cun}[2]{	\node at (#1,#2) [rectangle,inner sep=0pt,minimum size=\bsize] {$C_1$};}
\newcommand{\cdeux}[2]{	\node at (#1,#2) [rectangle,inner sep=0pt,minimum size=\bsize] {$C_2$};}
\newcommand{\ctrois}[2]{	\node at (#1,#2) [rectangle,inner sep=0pt,minimum size=\bsize] {$C_3$};}
\newcommand{\tts}[3]{	\node at (#1,#2) [rectangle,draw=black,fill=black!20,inner sep=0pt,minimum size=\bsize] {$#3$};}
\newcommand{\ttz}[2]{\tts{#1}{#2}{0}}
\newcommand{\ttu}[2]{\tts{#1}{#2}{1}}

\newcommand{\stt}[3]{	\node at (#1,#2) [rectangle,draw=black,fill=green!10,inner sep=0pt,minimum size=\bsize] {$#3$};}

\newcommand{\signalun}[3]{			
	\foreach \x in {1,...,#3} {\sun{\x +#1 -1}{\x +#2 -1}}
}
\newcommand{\signalunp}[3]{			
	\foreach \x in {1,...,#3} {\sunp{\x +#1 -1}{\x +#2 -1}}
}
\newcommand{\signaluns}[3]{			
	\foreach \x in {1,...,#3} {\suns{-\x +#1 +1}{+\x +#2 -1}}
}
\newcommand{\signaldeux}[3]{			
	\foreach \x in {1,...,#3} {\sdeux{\x +#1 -1}{+3*\x +#2 -3}}
}
\newcommand{\signaltrois}[3]{			
	\foreach \x in {1,...,#3} {\strois{-\x +#1 +1}{+3*\x +#2 -3}}
}
\newcommand{\signalquatre}[3]{			
	\foreach \x in {1,...,#3} {\squatre{-\x +#1 +1}{+\x +#2 -1}}
}
\newcommand{\signalquatrep}[3]{			
	\foreach \x in {1,...,#3} {\squatrep{\x +#1 -1}{\x +#2 -1}}
}
\newcommand{\signalcinq}[3]{			
	\foreach \x in {1,...,#3} {\scinq{-\x +#1 +1}{+\x +#2 -1}}
}
\newcommand{\signalsix}[3]{			
	\foreach \x in {1,...,#3} {\ssix{\x +#1 -1}{+3*\x +#2 -3}}
}
\newcommand{\signalr}[3]{			
	\foreach \x in {1,...,#3} {\sr{-\x +#1 +1}{+\x +#2 -1}}
}
\newcommand{\signaltt}[5]{			
	\foreach \x in {1,...,#3} {\stt{#4*\x +#1 -1}{+\x +#2 -1}{#5}}
}

\newcommand{\coldieze}[3]{			
	\foreach \x in {1,...,#3} {\dieze{#1}{\x +#2 -1}}
}	
\newcommand{\colM}[3]{			
	\foreach \x in {1,...,#3} {\sM{#1}{\x +#2 -1}}
}	
\newcommand{\colc}[3]{			
	\czero{#1}{#2}
	\foreach \x in {3,...,#3} {\cun{#1}{\x +#2 -2}}
}	
\newcommand{\colcd}[3]{			
	\cdeux{#1}{#2}
	\foreach \x in {3,...,#3} {\ctrois{#1}{\x +#2 -2}}
}	

%

In the following, we describe such a $\pset$-universal CA $\ACU$ with radius $1$ and alphabet $\alphU$.

We denote by $\pseto$ the set of CA of $\pset$ with radius $1$ and alphabet of size $2^\p$ for some $\p\in\NN$. 
One may easily describe for any CA $\ACB\in\pset$ a CA $\ACA\in\pseto$ such that $\ACB\fsimu\ACA$.
Using transitivity of $\fsimu$, it will be sufficient, in order to prove $\pset$-universality of $\ACU$, to exhibit for any CA $\ACA\in \pseto$ an onto local map $\fmap_\ACA$ from $\csetU$ to $\csetA$ and an integer $\tau_\ACA$ such that $\ACU^{\tau_\ACA} \circ \fmap_\ACA = \fmap_\ACA \circ \ACA$.

To do so, for each $\ACA$, we introduce an integer $l_\ACA$ and a dichotomy on words of $\alphU^{l_\ACA}$.
\begin{itemize}
\item on the one side we have what we call \emph{$\ACA$-correct macrocells} (or $\ACA$-macrocells). They encode information about a current state $x\in\alphA$, about the local rule of $\ACA$, and a machinery used to apply this rule to update the current state. In almost any case, they will be interpreted through $\fmap_\ACA$ as $x$.
\item on the other side we call all the other patterns \emph{$\ACA$-incorrect}, and they will be interpreted as the persistent state of $\ACA$.
\end{itemize}

The idea behind the local rule of $\ACU$ is to make every $\ACA$-macrocell determine if
it is surrounded by other $\ACA$-macrocells. If this is the case, then
interaction is possible, and the current state of the neighbor will be taken into account to compute the new current state, following the rule of $\ACA$. Otherwise, there is no interaction, the $\ACA$-macrocell evolves considering every $\ACA$-incorrect neighborhood as
a persistent state neighbor. The difficulty is that although
correctness is related to the particular CA being simulated, every
configuration must evolve correctly for every possible CA.

The proof of universality uses the combination of two key properties:
on one hand, correct patterns remain correct and evolve according to
the rule being simulated, even if not surrounded by correct patterns
(lemma \ref{correctmacrocells}); on the other hand, incorrect
patterns are interpreted as the persistent state and never become
correct (lemma \ref{incorrectmacrocells}).

To make the construction of $\ACU$ readable, we describe its state set
as a superposition of several layers: the main layer $M$
contains most of the information about the simulation and the macrocells informations; \emph{signals}
layers are used to manage the evolution of the main layer; and
\emph{clock} layers guarantee synchronizations.

%
%
%

\subsubsection{Correct macrocells description}

In the following we consider a simulated CA $\ACA\in\pseto$ with radius $1$ and state set $\alphA$ of size $n=2^\p$. We use a canonical binary enumeration of the state set, in which the first word ($0^{log(n)}$) represents a persistent state of $\ACA$, denoted $p_\ACA$.
Our $\ACA$-correct macrocells will be words of length $l_\ACA$ (specified above) whose main layer follow the pattern
\[ \#\ C_i\ |Transition\ table | \ |State| \  |memory|\ \#\]

\begin{itemize}
\item $\#$ are delimiters, they never appear or disappear during the computation
\item $C_i$ is the \emph{control state} used to control the successive steps of computation.
\item $|Transition\ table |$ is the binary description of the transition table of $\ACA$.
\item $|State|$ contains the binary value of the current state of the macrocell.
\item $|memory|$ is a binary area which will be used to keep the values of the neighbors' current states before computing the new current state of the macrocell.
\end{itemize}

$|Transition\ table |$, $|State|$ and $|memory|$ are encoded with disjoint binary sub-alphabets.
A cell whose state belongs to one of those sub-alphabets will never change sub-alphabet.
Moreover, the states of the transition table's cells are never modified.

The current state description is ${log(n)}$-bits-long. In the transition table, images are ordered canonically, so the length of the description is simply $n^3log(n)$. The memory should be at least  $2{log(n)}$-bits-long in order to contain current state values of the two neighbors. 
But in order to simplify some proofs, we chose $l_\ACA$ such that it is at least half the total size of the macrocell, and such that the function $\ACA\rightarrow l_\ACA$ is one-to-one.

Most of the computation will happen on those very constrained patterns. 
In the next definition, we add an extra constraint on the control state to obtain $\ACA$-correct macrocells.

By stability of the sub-alphabets, such a correct $\ACA$-macrocell remains correct, but the state value may change. This is why we take some care when defining the current state value associated to the macrocell.


\begin{defi}\label{coAdef}
A word $u\in\alphU^{l_\ACA}$ of length $l_{\ACA}$ is said to be \emph{a $\ACA$-correct macrocell}s, denoted by $u\in \coA$, if its main layer follows the structure defined above (correct sub-alphabets for each cell, and correct transition table of $\ACA$), and if its control state is in $C_0$.

For each such $\ACA$-correct macrocell $u$, we define its \emph{associated state value} $v(u)\in\alphA$, which is the state described by its current state value after $l_\ACA$ steps of computation by $\ACU$. And this state value $v(u)$ only depends on $u$.
\end{defi}

To ensure that the sate value only depends on $u$, the memory area is used as a buffer to prevent modification of the current state value coming from the left, before the computation has been initialized. Details will be given in the following.

By extension we may sometime call $\ACA$-macrocells words following the general pattern, even with non-$C_0$ control state, in particular when they are images of a $\ACA$-correct macrocell.

\subsubsection{The local rule.}

We describe the local behavior of $\ACU$ starting from a correct $\ACA$-macrocell.
The local rule will first determine which neighbors it may interact with (\textit{Check of the neighbor's length and synchronization}, and \textit{Transition table and state encoding check}), and then compute its new current state according to the rule of $\ACA$ and eventually the value of those neighbors (\textit{New current state computation}).

In order to guarantee the synchronization between $\ACA$-macrocells, we specify the duration of each step, and even of some sub-steps. It is done by a clock, which use specific layers of the states; their existence is proved by the following lemma:

\begin{lemm}\label{clock}
For any $k,h\in \NN \setminus \{1\}$, 
there exist a CA, and two states $q_s$, and $q_f$ such that the leftmost cell of an area delimited by two $\#$ separated by $l-2$ cells turns to state $q_f$ at some time $t>k\cdot{}l^2+h\cdot{}l$ iff this cell was in state $q_s$ exactly $k\cdot{}l^2+h\cdot{}l$ steps before.
Moreover, this property is guaranteed independently of what is outside the two $\#$.
\end{lemm}

At the beginning of each step, the control state $C_i$ will turn to $C_{i+1}$, initiate the corresponding clock, and initiate some signals which will manage the evolution. Those signals are distinct states propagating on upper layers of the configuration, and interacting with the main layer and other signals.
We say that a signal \textit{belongs} to a macrocell if it was generated in this macrocell's area, between the two $\#$. And, thanks to our evolution rule, a signal always knows if it is in its cell or in the area to the right or left of its cell.
It is also useful sometimes to make signals carry one extra bit of information. It is simple to do it using distinct states, since the number of bits is bounded.

\begin{figure}[t]
\begin{minipage}[b]{0.49\linewidth}
\scalebox{0.6}{
\tiny
\begin{tikzpicture}
	\def\h{25}
	\def\hp{\h +0.5}
	\pgfsetxvec{\pgfpoint{.3cm}{0cm}}
	\pgfsetyvec{\pgfpoint{0cm}{.3 cm}}
	\draw[step=1] (-.5,0) grid (24.5,\hp);
	\pgftransformshift{\pgfpoint{0.15cm}{0.15cm}};

\coldieze{0}{0}{\h}
\coldieze{11}{0}{\h}
\coldieze{12}{0}{\h}
\coldieze{23}{0}{\h}
\colc{1}{0}{\h}
\colc{13}{0}{\h}
\cun{1}{\h-1}
\cun{13}{\h-1}
\signalsix{6}{6}{6}
\signalun{1}{1}{6}
\signalun{13}{1}{11}
\signalcinq{23}{12}{12}
\signalquatre{12}{1}{12}
\signalquatrep{1}{13}{11}
\srp{12}{24}

\end{tikzpicture}}
\end{minipage}
\hfill
\begin{minipage}[b]{0.49\linewidth}
\scalebox{0.6}{
\tiny
\begin{tikzpicture}
	\def\h{47}
	\def\hp{\h +0.5}
	\pgfsetxvec{\pgfpoint{.3cm}{0cm}}
	\pgfsetyvec{\pgfpoint{0cm}{.3 cm}}
	\draw[step=1] (-.5,0) grid (24.5,\hp);
	\pgftransformshift{\pgfpoint{0.15cm}{0.15cm}};
\coldieze{0}{0}{\h}
\coldieze{11}{0}{\h}
\coldieze{12}{0}{\h}
\coldieze{23}{0}{\h}
\colc{1}{0}{\h}
\colc{13}{0}{\h}
\cdeux{1}{\h-1}
\cdeux{13}{\h-1}
\signaldeux{0}{0}{12}
\signaldeux{12}{0}{12}
\signalun{0}{0}{12}
\signalun{13}{1}{11}
\signalunp{12}{12}{12}
\signaluns{23}{24}{12}
\signaltrois{17}{18}{6}
\signalr{12}{35}{11}
\dieze{0}{0}
\dieze{12}{0}
\dieze{11}{35}
\end{tikzpicture}
}
\end{minipage}
\caption{Successful left (resp. right) neighbor test by the right (resp. left) macrocell (mix of main and signal layers for easier reading)}
\label{fig:firststep}
\end{figure}
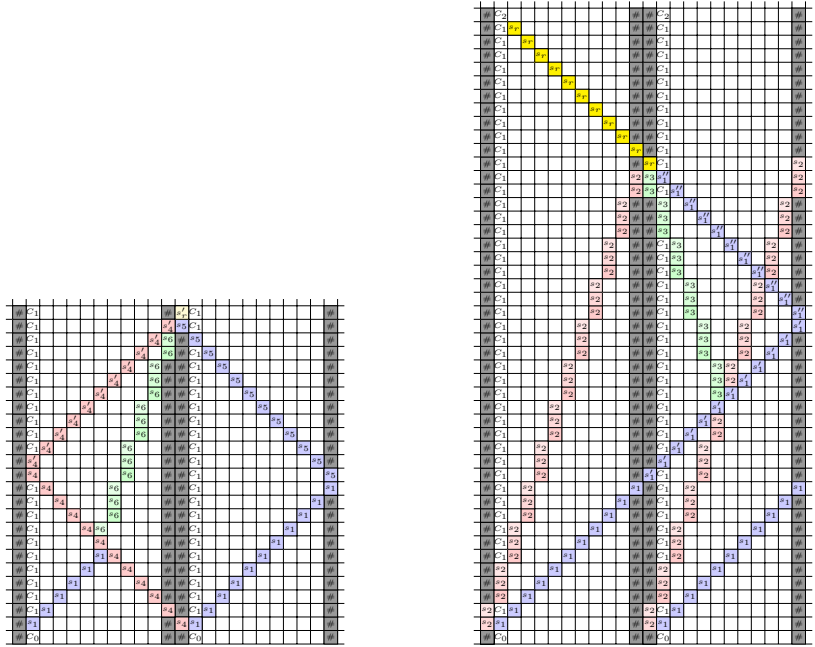

\paragraph{\textbf{Check of the neighbor's length and synchronization. } $C_0 \rightarrow C_1$} (that is to say that when the control cell's state is $C_0$ it becomes $C_1$): 

Recall that we are interested to the behavior in the case of an $\ACA$-correct macrocell.
When $C_0$ becomes $C_1$, it initializes two \emph{control bits} with value $0$, in the main layer of the control cell, and it launches signals. Since the construction is classical, we simply illustrate the desired behaviors by figure~\ref{fig:firststep}. Those two pictures illustrate the signal machinery in the case of respectively left and right neighbors of same length and with state $C_0$ appearing simultaneously (what we call synchronized).
Every transition whose image is one of the signal involved in this checking appears on those pictures.

The first signals $s_1$ and $s_4$ erase all signals belonging to our macrocell.
Together with the length of the memory being bigger than the length between the control state where signals are generated and the end of the current state area, it constitutes the protection of the current state value from eratic signals coming from outside the macrocell, and justifies that in definition \ref{coAdef} $v(u)$ is a function of $u$.

%
%
%
%
%

If the neighbors have same length and are synchronized, this whole step takes 4 times the length of the macrocell, $l_\ACA$. 
After $4\cdot{}l_\ACA$ steps, the control state $C_1$ becomes $C_2$, and if it did not receive a positive result from one side, it concludes that the involved neighbor is incorrect.
This is managed using a clock signal (with $h=2$ and $k=0$ in lemma \ref{clock}) initialized by $C_0$ on a specific layer. When $q_f$ is raised on this layer, $C_1$ becomes $C_2$. 
The important point is that we ensure the following property.
\begin{lemm}
The control state of an $\ACA$-correct macrocell becomes $C_2$ exactly $4\cdot{}l_\ACA$ steps after $C_0$ appeared. At this step each control bit has turned to $1$ iff the corresponding neighboring macrocell has same length and synchronisation than the considered macrocell.
\end{lemm}
The proof of this lemma is direct for the length but asks to enter into some more (simple but fastidious) details for the synchronization part.


\paragraph{ \textbf{Transition table and state encoding check}. $C_2 \rightarrow C_3$} :

In this step, for each neighboring pattern with same length and synchronization, the macrocell checks whether the transition table and the current state are compatible with its own (same lengths, and same content for the transition table) or not.

When $C_2$ appears it launches the following test for each neighbor whose corresponding control bit was $1$, and initializes two fresh bits to $0$.
First, a signal is generated and puts a \emph{mark} (that is to say a non-moving signal) on the first cell of the transition table of its macrocell, and another mark on the first cell of the transition table of the neighbor it checks. Then signals are exchanged between those two marks that will each time carry the binary state of the cell pointed by one mark to the next unchecked cell of the other macrocell; it compares this cell's binary state to the carried binary state, and push the mark by two cells. If no difference is detected and if both marks reach the end of the transition tables simultaneously, a correctness signal is sent to the control state. 

After the transition table has been checked, the same mechanism is used to check that the current state encoding areas have same length.
%
%
%
At the end of those tests, the results are sent to the control cell which again keeps the information on two control bits. 
%
For each cell of the transition table or the current state, checking takes $2\cdot{}l_\ACA$ steps. So checking a whole neighbor takes less than $2\cdot{}l_\ACA^2$.
Again, a clock is used to make this test last exactly $2\cdot{}l_\ACA^2$ steps. Then the control cell is turned to $C_4$.

\begin{lemm}
The control state of our macrocell becomes $C_4$ exactly $2\cdot{}l_\ACA^2$ steps after $C_2$ appeared. At this step each control bit has turned to $1$ iff the corresponding neighboring pattern has same length, synchronization, and if the lengths of the transition tables and current states, and the content of the transition table are equal. In this case we say that this pattern is \emph{compatible} with our $\ACA$-macrocell.
\end{lemm}

The proof of this lemma is straightforward. Keep in mind that some signals erased all erratic signals that could interact with our cell at a previous step.



\paragraph{\textbf{New current state computation}.} $C_4 \rightarrow C_5$ :

After all the tests have been done, the new current state has to be computed.
We need to explicit how we consider the neighboring pattern.
In the following, what we call \emph{detected state} of one such pattern by our macrocell will be: either the persistent state if the neighbor is non-compatible with our $\ACA$-macrocell, or its current state if this is a compatible $\ACA$-macrocell.

At first, the detected states of the left and right neighbors are written to the memory. It is written in the binary memory alphabet. Each detected state is written on $log(n)$ cells. If one neighbor is compatible, we copy its current state value to the memory using marks and signals similarly to the previous step. If it is not compatible, we write $0^{log(n)}$, the length being the same as that of the current state area.
We add a clock to specify that copies last exactly $2\cdot{}l_\ACA^2$ steps,
the neighbor being correct or not.

After $2\cdot{}l_\ACA^2$ additional steps, the search for the image in the transition table starts. It consists in reading the binary word formed by the three image states (the current state of the cell followed by the two detected states copied in the memory), and turning it into a unary position in the transition table. It is then possible to place a mark at this position, and finally copy this pointed state to the current state area.
We make the reading of the position last $4\cdot{}l_\ACA^2$ steps. 
And copying the new state lasts $2\cdot{}l_\ACA^2$. 
After the whole computation step, which lasts $7\cdot{}l_\ACA^2$, the control state turns to $C_5$.

Finally one step of simulation is completed after exactly $\tau_\ACA=9\cdot{}l_\ACA^2+4\cdot{}l_\ACA$ steps. After this time the control state turns to $C_5$.

To become $C_0$ again, and launch a new step of computation, we add another condition. We ask a clock launched exactly $\tau_\ACA$ steps before to raise a flag.
And obviously this clock may only be launched by $C_0$.
It is realized using again signals of the lemma \ref{clock} computing on one more layer.

The state set of the universal CA is given by $\alphU = M \times S
\times C \cup \{C_f\}$ with
\begin{itemize}
\item the main layer : 
	$ M = \{C_0,C_5\}\, \cup\, \{C_i\}_{i\in\{1,..,4\}}\times\{0,1\}^2 
				\,\cup\, \{0_{i},1_{i}\}_{i\in \{tt,cs,m\}}$
\item the signals layers :
 	$ S =\ $\Large $\times$\normalsize$_{i\in I} \{s_i\}\ \times\ $\Large $\times$\normalsize$_{j\in J} (\{s_j\}\times \{0,1\})$
\item the clocks layers (see lemma \ref{clock}), one for each duration needed.
	$ C = (\{0,1\}\times \{s_i\}_{i\in I_c})^4$
\item $C_f$ is a single persistent state ensuring that $\ACU\in\pset$
\end{itemize}

Yet, the transition rule of $\ACU$ is partially specified, we call correct transitions those defined up to now, in the case of correct macrocells. But the other transitions may not be chosen arbitrarily. We specify the following behaviors:
\begin{itemize}
\item $C_f$ is never modified by any transition
\item the main layer is never modified by a non-correct transition: they act as the identity on the main layer.
\item concerning the signal layer, apart from the collisions corresponding to the behavior described in the previous steps, all signals may cross each other (each kind of signal is evolving on its own layer). However, except for transitions involved in the behavior described above, any signal that crosses a $\#$ is destroyed.
\end{itemize}

\subsubsection{Interpretation}
We now describe the continuous onto map $\fmap_{\ACA}:\csetU \rightarrow \csetA $ associated to $\ACA$. 
This map is induced by a local map $f_\ACA$ from patterns of shape $l_{\ACA}$ to individual states of $\ACA$. More precisely, using notation from proposition~\ref{prop:hedlund}, we have ${r=0}$, ${z_1=l_{\ACA}}$, ${z_2=1}$, ${t_2=1}$ and ${t_1=\tau_{\ACA}}$.

If $p_\ACA$ is the persistent state of $\ACA$, the local map $\psi_\ACA$ is defined as follows:
\begin{enumerate}
\item $\forall u\not\in\coA, f_\ACA(u)=p_\ACA$

\item $\forall u\in\coA, f_\ACA(u)=v(u)$, with $v(u)$ the value from definition \ref{coAdef}


\end{enumerate}

\subsubsection{Proof of theorem~\ref{theo:univpe}}
The proof of the theorem relies on the two following lemmas. They 
are consequences of the construction, the intermediate lemmas and the clock lemma.

\begin{lemm}\label{correctmacrocells}
$\forall$ $c\in\csetU$, $t_0\in\NN$, if ${\ACU^{t_0}(c)_{[0,l_{\ACA}-1]}\in\coA}$, then ${v = \ACU^{t_0+\tau_\ACA}(c)_{[0,l_{\ACA}-1]}\in\coA}$, and ${\psi_\ACA(v)=
\locA(\psi_\ACA(c_{[-l_{\ACA},-1]}),\psi_\ACA(c_{[0,l_{\ACA}-1]}),\psi_\ACA(c_{[l_\ACA,2\cdot{}l_{\ACA}-1]}))}$.
\end{lemm}


\begin{lemm}\label{incorrectmacrocells}
If $\exists$ $t \geq \tau_{\ACA}$, $c \in \csetU$ such that $u=\ACU^t(c)_{[0,l_{\ACA}-1 ]}\in \coA$ then $v=\ACU^{t-\tau_{\ACA}}(c)_{[0,l_{\ACA}-1 ]}\in \coA$.
\end{lemm}




We can finally prove our main claim: $\forall \ACA\in \pseto$,
$\ACA\fsimu\ACU$.  We use the characterization of
proposition~\ref{prop:hedlund}. Let $\ACA\in\pseto$ the associated length $l_{\ACA}$ and function $\fmap_{\ACA}$ are
defined as explained before. First, $\fmap_\ACA$ is local (by
definition) and onto, because correct macrocells are enough to encode
any state of $\ACA$ and thus concatenations of correct macrocells
allow to encode any configuration of $\ACA$. Second, we have
$\fmap_\ACA\circ\ACU^{\tau_\ACA}=\ACA\circ\fmap_\ACA$. To see this we
discuss on the pattern of shape $\rect{l_\ACA}$ at position $0$ and
the rest follows by translation. If this pattern is not in $\coA$ its
image after $\tau_\ACA$ steps remains out of $\coA$ (lemma
\ref{incorrectmacrocells}). If conversely this central word belongs to
$\coA$, lemma \ref{correctmacrocells} gives the desired property.










\section{Perspectives}

A natural extension of our work could be to generalize the
construction to cellular automata having an equicontinuous point. The
idea would be to use blocking words as a replacement for the
persistent state. But it seems much harder, if not impossible.

Besides, the main open question left by this paper is the existence
of universal CA. We conjecture that they do not exist and more
precisely that no CA can simulate all products of shifts. A possible
way to obtain this negative result would be to study limit sets: by a
compacity argument, one can show that a universal CA must have a
universal limit set. The main obstacle is that subshifts that are
limit sets of CA are not well characterized.  

Finally, we also leave open the existence of universal SFT and
universal surjective CA in dimension 1.

\bibliographystyle{plain}
\bibliography{ac}

\newpage
\appendix

\section{Proofs from section~\ref{sec:univdef}}

\begin{proof}[Proposition~\ref{prop:hedlund}]
  First, an onto local map from $\Sigma_1$ to $\Sigma_2$ with shapes
  $\rect{z_1}$ and $\rect{z_2}$ induces a factoring relation from the
  $\mo$-scaled action of $\Sigma_1$ onto the $\mo'$-scaled action of
  $\Sigma_2$ with
  \[\mo=(\vec{z_1})_1\ZZ\times\cdots (\vec{z_1})_d\ZZ\]
    and
  \[\mo'=(\vec{z_2})_1\ZZ\times\cdots (\vec{z_2})_d\ZZ.\]

    Conversely, suppose that the relation $\Sigma_1\fsimu\Sigma_2$ is
    realized by a factor map $\pi$ from the $\mo$-scaled action of
    $\Sigma_1$ onto the $\mo'$-scaled action of $\Sigma_2$ with
    \[\mo=(\vec{z_1})_1\ZZ\times\cdots (\vec{z_1})_d\ZZ\]
    and
    \[\mo'=(\vec{z_2})_1\ZZ\times\cdots (\vec{z_2})_d\ZZ.\]
    Consider now each pattern $p\in Q_2^{\rect{z_2}}$. Since the
    cylinder $C_p$ defined by 
    \[C_p = \{c\in Q_2^{\ZZ^d} : \patternat{z_2}{0}{c}=p\}\] is both
    open and closed, so is $\pi^{-1}(C_p)$. By compacity, and since
    cylinders form a basis of the topology, we get that
    $\pi^{-1}(C_p)$ is a finite union of cylinders of
    $Q_1^{\ZZ^d}$. We can suppose without loss of generality that they
    are all of shape $\rect{(2r+1)z_1}$ for some large enough $r$
    (finite unions of cylinders of small shape can always be defined
    as finite unions of cylinders of larger shapes). Doing this with
    the same value of $r$ for all $p$, we get a (possibly partial)
    function $f$ from $Q_1^{\rect{(2r+1)z_1}}$ to
    $Q_2^{\rect{z_2}}$. By eventually completing $f$ and by definition
    of the factoring $\pi$ between $\mo$ and $\mo'$-scaled actions,
    $f$ induces a local map from $Q_1^{\ZZ^d}$ to $Q_2^{\ZZ}$
    associated with shapes $\rect{z_1}$ and $\rect{z_2}$. It is onto
    because $\pi$ is onto.

    For cellular automata, the reasoning is similar and adding the
    temporal component in actions translates exactly into the desired
    property of weak commutation between the global maps of cellular
    automata and the onto map between configuration spaces.\qed
\end{proof}

\begin{proof}[Theorem~\ref{theo:nouniv}]
  For the case of surjective CA, it is enough to notice that
  surjectivity is preserved by the relation $\fsimu$. Indeed, if
  $F\fsimu G$ we have
  \[\phi\circ G^{t_1} = F^{t_2}\circ\phi\] for some onto map $\phi$. Therefore
  $F$ must be surjective if $G$ is surjective.
  
  Then the proof follows from Kari's theorem \cite{kari94} establishing
  that surjective CA are not recursively enumerable.  Indeed, given a
  surjective universal CA $U$, we could enumerate thanks to the local
  presentation of factors (proposition~ \ref{prop:hedlund}) all CA F
  such that $F\fsimu U$: they are all surjective (surjectivity is
  preserved by factor) and all surjective CA are among them
  (universality).\\

  We consider now the case of subshifts of finite type.  Without loss
  of generality, any subshift of finite type can be presented as a
  subshift $\Sigma_L$ where $L$ is a finite set of patterns having all
  the same shape $\rect{z}$ for some $z$. By Berger's
  theorem \cite{berger} the set of such $L$ verifying that $\Sigma_L$
  is not empty can not be recursively enumerated. We show below that
  the existence of a universal subshift of finite type implies the
  existence of an algorithm of enumeration of all $L$ of the form
  above such that $\Sigma_L$ is not empty.

  So suppose that there exists some universal subshift of finite type
  $\Sigma_{L_U}$ where $L_U$ is a set of $Q_U$-patterns of shape
  $\rect{z_U}$. Obviously, $\Sigma_{L_U}$ must be non-empty. For any
  $L$ and any pattern $p$ of larger shape, we say that $p$ is
  $L$-valid if it contains no occurrence of any pattern from $L$
  (occurrence requires that one shape is completely included into the
  other).
  
  Let $L$ be a set of $Q$-patterns of shape $\rect{z}$ and $\psi$ be a
  local map from $Q_U^{\ZZ^d}$ to $Q^{\ZZ^d}$ associated to shapes
  $\rect{z_1}$ and $\rect{z_2}$. Consider the minimal shape
  $\rect{z_+}$ containing both $\rect{z}$ and $\rect{2z_2}$. Since
  $\psi$ is local, one can check in finite time the following property
  called \emph{validity property}: any pattern $p$ of shape
  $\rect{z_+}$ which has a $L_U$-valid preimage via $\psi$ is
  $L$-valid (the size of preimages of finite patterns depends on the
  radius $r$ associated to $\psi$ but details don't matter here). By
  the definition of local maps and the hypothesis on shapes, this
  property implies that $\psi(\Sigma_{L_U})\subseteq\Sigma_L$ and
  therefore $\Sigma_L\not=\emptyset$ (the choice of shape
  $\rect{2z_2}$ ensures that validity is checked inside blocks of shape
  $\rect{z_2}$ but also across the boundary between two such adjacent
  blocks).

  It follows that we can recursively enumerate couples $(L,\psi)$
  having the property above. More precisely, maps $\psi$ are
  enumerated via their local presentation (shapes, radius and local
  function). This way, we can enumerate a list of finite languages $L$
  such that $\Sigma_L$ is not empty. To conclude the proof it is
  sufficient to show that all $L$ such that $\Sigma_L\not=\emptyset$
  are present in the list. Suppose by contradiction that some $L$ over
  alphabet $Q$ with $\Sigma_L\not=\emptyset$ is such that no local map
  from $Q_U^{\ZZ^d}$ to $Q^{\ZZ^d}$ verifies the validity property
  above. By universality of $\Sigma_{L_U}$, there exists a local map
  $\psi$ sending $\Sigma_{L_U}$ to $\Sigma_L$. Let $r$, $\rect{z_1}$
  and $\rect{z_2}$, and local function $f$, be the parameters
  associated to $\psi$. For any $k\geq 0$ we can define the same map
  $\psi$ with another presentation by increasing artificially the
  radius $r$ to $kr$ and changing the local function $f$ accordingly
  (shapes are kept unchanged). We call it the k$^{th}$ presentation of
  $\psi$. Since, by hypothesis on $L$, no such presentation has the
  validity property, we deduce that there must exist some finite
  pattern $p$ which is not $L$-valid and such that, for any $k$, $p$
  has a $L_U$-valid preimage under the k$^{th}$ presentation of
  $\psi$. Therefore, by a simple compacity argument, there exists
  $c\in\Sigma_{L_U}$ such that $\psi(c)$ has an occurrence of
  $p$. Hence, $\psi(c)\not\in\Sigma_L$ which is a contradiction.\qed
\end{proof}

\section{Proofs from section~\ref{sec:univpe}}

\begin{proo}[of lemma \ref{clock}]
We first build a CA that satisfies our lemma for $k=2$, $h=0$.
Its state set will be made of one binary layer, and a signal layer.
The behavior is simple: when $q_s$ appears it generates a signal that will keep oscillating between the $\#$. When the signal is generated for the first time, it initialize the area, turning the first binary cell to $1$ and the other one to $0$s. Then, the signal keep moving from right to left and back between the $\#$. Each time it goes to the right, it turns one more binary cell to $1$. And when the rightmost cell's binary layer is finally turned to $1$ a new special signal is sent to the left which will generate the $q_f$.

If two or more signals crosses, one of them may survive. If one of them is initializing it will survive.

So, in $2.l$ steps of computation, the total number of $1$s may be non increasing only in the following cases:
\begin{itemize}
\item if there is no signal at all
\item if all cell's binary layer is already $1$ and in this case a $q_f$ was generated
\item if an initialization signal has been sent.
\end{itemize}

In particular, if a $q_f$ appears at some step, then in the previous $4.l$ steps, a $1$ was generated. And in each previous $2.l$ step, at least a $1$ was generated. 

Thus, in the previous $2.l^2$ steps, at least one initialization signal was launched and a $q_s$ has appeared. But by construction, after a $q_s$ state appears, the first $q_f$ state appears only exactly $2.l^2$ steps later.

It concludes the proof of the clock lemma in case $k=2$, $h=0$. For other values of $k$, simply slow down the signal going right to left. For other values of $h$, after the end of the quadratic part, launch a signal that will go right with speed $1$ and come back left with speed $1/(h-1)$ before raising $q_f$.
\end{proo}


\begin{proo}[of lemma \ref{incorrectmacrocells}]

By definition of a correct pattern, $u$ is given by:
$$ \#\ C_0\ |Transition\ table | \ |State| \  |memory|\ \#$$

First of all, the $\#$ are never created or destroyed.
The transition table and maximal state information are never modified, so they are the same in $u$ and in $v$.
And the sub-alphabet corresponding to control state, current state and memory alphabets are stable. The structure of $v$ is the same as this of $u$.
To prove our lemma it remains to prove that the second letter in $v$ is $C_0$, and that the current state value is smaller than the maximal value.

But, to make $C_0$ appear, at step $t$, a  signal $q_f$ was raised by the global clock, which implied, using the clock lemma, that the second letter of $v$ is $C_0$.

Thus all tests are launched. 
\end{proo}

\end{document}